\documentclass[12pt, a4]{article}
\usepackage{sbc-template}
\usepackage{graphicx,url}
\usepackage{color}

\usepackage[english]{babel}   
\usepackage[utf8]{inputenc}

\title{A Discussion about Computational Challenges of Programmable
Money in Blockchain-based CBDCs} 

\author{Arlindo F. da Conceição\inst{1}, Roman Vitenberg\inst{2}}

\address{Universidade Federal de São Paulo (UNIFESP)
\nextinstitute
Blockchain Lab, University of Oslo (UiO)\\
\email{arlindo.conceicao@unifesp.br, romanvi@ifi.uio.no}
}

\begin{document} 

\maketitle

\begin{abstract}
This article discusses the implementation of programmable money on DLT-based CBDCs. After briefly introducing what programmable money is, we enumerate some initiatives worldwide and discuss the critical steps for implementation. We look at the challenges from the Computer Science perspective. Four aspects were analyzed: architectural design, security, scalability, and energy consumption. 
\end{abstract}


\section{Introduction}


Programmable money refers to the capacity of a digital currency to be programmed to impose specific rules and conditions for its use. This is typically done using smart contracts over blockchain networks. 
The money is a token in a blockchain network, and the smart contracts can automatically execute transactions based on certain pre-defined conditions. For example, a smart contract could be set up to deliver a good only after payment confirmation. Another example is the automatic interest payment on a loan based on a pre-determined interest rate and payment schedule.


Programmable money can reduce transaction costs and increase efficiency by automating financial processes~\cite{sethaput2023blockchain}. It also opens up new possibilities for financial innovation, such as disintermediation, developing new social policies, and fostering circular economy. By reducing human mediation and costs, programmable money can make machine-to-machine transactions and micropayments feasible.

Programmable money can potentially revolutionize the economy, making it ``smarter'',  but its implementation is far from trivial. This short article explores the computer science issues for programmable money implementation; we believe that design decisions can impact security, scalability, and usability, having a fundamental impact on the solution's success. Despite their importance, issues not directly related to computer science, such as innovative business models, law regulations, social effects, tokenomics, and economic stability, are kept out of the scope.




\section{Literature about Blockchain-based Programmable CBDCs}

Several countries are testing CBDC solutions and have recognized the potential of programmable money. China, Singapore, European Union (Digital Euro), United States, and Sweden are some examples~\cite{tracker}.

The technology is in its infancy, but we can state that the software architecture of a Blockchain-based Programmable CBDC is described in the literature with up to four layers:

\begin{itemize}
    \item Layer 1, \textbf{tokens management}. It consists of a blockchain infrastructure, usually private in the case of CBDC, that is responsible for storing tokens and avoiding double-spending.
    
    \item Layer 2, \textbf{smart contracts execution}. Smart Contracts operate the transactions and, usually, are responsible for implementing the core of the business logic. This level can include the application programming interface (APIs), which specifies the functions that can be accessed by the users.
    
    \item Layer 3, \textbf{applications}. It is  usually a web application. It calls Smart Contracts and can implement a part of the  business logic.
    
    \item Layer 4, \textbf{wallets}. The user interface is a piece of software that manages the user's keys and access the systems. The wallets can also be responsible for part of the business logic.
\end{itemize}

Innovative use cases can be implemented using the infrastructure for Programmable Money~\cite{weber2022programmable}. The implementation, of course, can vary a lot. The consensus model (PoW, PoS, etc.), transaction model (UTXO ou account balance), token standard (for example, ERC20), and transaction fees are some of the variables with impact on security, usability, and scalability.

Recently, the \textit{Banco Central do Brasil} (BCB) published guidelines for the implementation of Digital Real~\cite{diretrizes}. Programmable money are not necessarily based on Blockchain, but the first directive defines which smart contracts will be used. The directive eighth specifies that the solution will be based on Distributed Ledgers. Last but not least, directive 9 expresses cybersecurity concerns. We can infer that the solution will be an implementation of a private Blockchain.

We use the Norwegian case as a baseline for this discussion. The central bank of Norway is testing a private Blockchain network based on the Hyperledger Besu technology~\cite{norway}. In this experiment, the money is an ERC20 token, and the programmability feature should be implemented using Solidity smart contracts.


\section{Computational Challenges}

This section enumerates some of the most critical challenges for programmable money implementation from the perspective of Computer Science.

\subsection{Architectural Design}

The first question is the architecture, which must consider the use cases and its requirements. Here, we considered only blockchain-based architectures. The solution must find trade-offs between scalability, decentralization, performance, transparency, privacy, usability, security, etc.

A good design should organize the solution in layers~\cite{jin2022cev}, detaching the components if possible. The architecture can follow the pattern of two layers~\cite{gudgeon2020sok}: one for transactions, the other for business implementation. 

The first layer comprises a network of \textbf{Validators} nodes, that are responsible for keeping the chain of blocks consistent. In this layer, choosing the correct consensus algorithm is essential. 
 The literature shows that the consensus algorithm usually is the bottleneck of the solution~\cite{androulaki2018hyperledger}.  CBDCs usually have a trusted party --- the Central Bank --- so they can use fast consensus algorithms, like RAFT. Other techniques such as sharding and geographical division can also be applied at this layer to improve the performance. The security requirements for this layer are critical.

The second layer can contain business logic implementation and additional services. It can offer, for example, the execution environment for smart contracts. It defines the API for the applications and must deal with high complexity integration process. This level must specify the token technology, the default choice is to use a fungible standard. But some CBDCs can also implement support for non-fungible tokens, that can be used to represent digital assets.




    



\subsubsection{Rules of representation}
The programmable money rules can be represented in different styles. The two most common representations are: \textbf{list of predicates} and \textbf{scripts}.

 A list of predicates is a technique similar to a firewall's configuration. The user describes a set of rules that have to be true to transaction confirmation. In the case of CBDC, the rules must be true to the transaction to be considered valid and stored on the ledger.

A CBDC based on smart contracts is based on scripts. The scripts are programs written in a programming language that defines the behavior of the money.

A list of predicates is more simple to implement but is less powerful. Scripting is a more powerfull technique and can be applied in more scenarios, but its complexity brings performance and vulnerability issues.

The representation model depends on the size and complexity of the list of programmable attributes. The behavior of a currency can be programmed in terms of several variables, for example: time, region or location, origin of funds, destiny, etc. Even off-chain variables could be considered. 

The market seems to prefer the flexibility of scripting models. One possible strategy is to support Turing-complete environments from the beginning, like EVMs, but to allow only limited execution capacity depending on use cases and security risks. The capacity could be expanded according to network maturity and continuous risk assessment~\cite{jansen2020smart}.

\subsection{Security}

Security is a critical concern for any CBDC. Security failures can diminish the public trust in the CBDCs~\cite{hansen2022security}.

Blockchain-based CBDCs have reduced risks of fraud related to centralization; on the other hand, they have opened new possible attacks due to decentralization. And it is still subject to well-known attacks, such as DDOS.

\subsubsection{Loss of keys}

One common problem with all digital currencies is the loss of private keys~\cite{Bonneau2015Sok}. Private keys are passwords that allow access to digital currency. A Central Bank is expected to offer a legal way to recover money access to the CBDC in case of key loss~\cite{kahn2020security}. This problem is specially hard if the CBDC allows anonymous users.

\subsubsection{Privacy}

Central banks have to safeguard users' data privacy. The privacy-preserving strategies may include strong encryption and robust regulatory frameworks.
Research in this area is still ongoing~\cite{lee2021survey}. Data anonymization, zero-knowledge proofs~\cite{takaragi2023secure}, and ring signatures~\cite{li2020blockchain} are some techniques explored in the literature.

\subsubsection{Correctness of smart contracts} 

The question here is how to assure that a smart contract is correct. It is especially important because programmable money deal with real monetary values. 

There is a vast literature about vulnerability detection in solidity smart contracts~\cite{de2021vulnerabilities}. 
However,
in the context of programmable money, the verification tools should be extended to check if a contract achieves the proposed objective. For example, suppose that a token was created to be spent only on organic food. But, on the next level of the supply chain, the money can be spent on fossil fuels and unsustainable production methods. It is an open problem that has to be addressed in the future.



\subsection{Scalability}

Scalability is the capacity to handle a high volume of transactions efficiently and effectively. In the CBDC context, some techniques to improve scalability are horizontal and vertical scaling, reducing transaction complexity, optimizing network infrastructure, sharding, off-chain processing, etc.

In the future, with more complex programmable money, scalability can be a massive challenge for big economies. It is worth mentioning that machine-to-machine and micropayment transactions can make scalability even more challenging.




\subsection{Energy consumption}

Finally, efficient energy consumption is a fundamental requirement. In a blockchain-based CBDC, every smart contract must be executed and validated in each validator node, multiplying the energy consumption impact. 
If a country successfully implemented a CBDC, the global impact of 1\% energy consumption in a CBDC implementation could be huge.



\section{Conclusion}

Programmable money has the potential to revolutionize the economy. But the technology is still in its early stages. In this article, we discussed a list of challenges that deserve attention and must be addressed by the computer science community. The list, of course, is incomplete. 

In the future, we will explore the innovative potential of specific use cases. We plan also to investigate methods to formally represent the global goals of programmable money, and how to evaluate if the goals were achieved.


\bibliographystyle{sbc}
\bibliography{sbc-template}

\end{document}